\documentclass{article}\usepackage[left=3cm, right=3cm, top=2cm, bottom = 2cm]{geometry}

\usepackage{graphicx}

\usepackage[xindy]{glossaries}

\newacronym{ltl}{LTL}{Linear-time Temporal Logic}

\newacronym{ctl}{CTL}{Computation Tree Logic}

\newacronym{ptl}{PTL}{Probabilistic Temporal Logic}

\newacronym{pctl}{PCTL}{Probabilistic Computation Tree Logic}

\newacronym{pctl*}{PCTL*}{Probabilistic Computation Tree Logic*}

\newacronym{tctl}{TCTL}{Timed Computation Tree Logic}

\newacronym{stl}{STL}{Signal Temporal Logic}

\newacronym{ltl-x}{LTL-X}{a version of Linear Time Logic without the `next' operator}

\newacronym{mtl}{MTL}{Metric Temporal Logic}

\newacronym{iactlk-x}{IACTLK-X}{a fragment of the temporal-epistemic logic CTLK, without the `next' operator}

\newacronym{mucalc}{$\mu$-calculus}{a strict superset of other temporal logics}

\newacronym{fotl}{FOTL}{First-Order Temporal Logic}

\newacronym{bdi}{BDI}{Belief-Desire-Intention}

\newacronym{prs}{PRS}{Procedural Reasoning System}

\newacronym{are}{ARE}{Autonomous Reasoning Engine}

\newacronym{dmars}{dMARS}{distributed Multi-Agent Reasoning System}

\newacronym{fsm}{FSM}{Finite-State Machine}

\newacronym{pfsm}{PFSM}{Probabilistic Finite-State Machine}

\newacronym{apfsm}{APFSM}{\textit{Autonomous} Probabilistic Finite-State Machine}

\newacronym{pha}{PHA}{Probabilistic Hybrid Automata}

\newacronym{fsa}{FSA}{Finite-State Automata}

\newacronym{ta}{TA}{Timed Automata}

\newacronym{gspn}{GSPN}{Generalised Stochastic Petri Net}

\newacronym{mopn}{MOPN}{Marked Ordinary Petri Net}

\newacronym{ba}{BA}{B\"{u}chi Automata}

\newacronym{dtmc}{DTMC}{Discrete-Time Markov Chain}

\newacronym{pars}{PARS}{Process Algebra for Robot Schemas}

\newacronym{fsp}{FSP}{Finite State Processes}

\newacronym{piadl}{$\pi$ADL}{$\pi$ calculus combined with ADL}

\newacronym{csp}{CSP}{Communicating Sequential Processes}

\newacronym{ccs}{CCS}{Calculus of Communicating Systems}

\newacronym{wsccs}{WSCCS}{Weighted Synchronous CCS}

\newacronym{csp|b}{CSP$\|$B}{an integrated formalism combining CSP and B}

\newacronym{fdr}{FDR}{the Failures-Divergences Refinement checker}

\newacronym{jpf}{JPF}{Java PathFinder}

\newacronym{ajpf}{AJPF}{Agent Java PathFinder}

\newacronym{mcmas}{MCMAS}{Model Checker for Multi-Agent Systems}

\newacronym{ltlmop}{LTLMoP}{Linear Temporal Logic MissiOn Planning}

\newacronym{adl}{ADL}{Architecture Description Language}

\newacronym{aadl}{AADL}{Architecture Analysis and Design Language}

\newacronym{lts}{LTS}{Labelled-Transition System}

\newacronym{uml}{UML}{Unified Modelling Language}

\newacronym{sysml}{SySML}{Systems Modelling Language}

\newacronym{robotml}{RobotML}{Robot Modelling Language}

\newacronym{dsl}{DSL}{Domain Specific Language}

\newacronym{sct}{SCT}{Supervisory Control Theory}

\newacronym{ide}{IDE}{Integrated Development Environment}

\newacronym{mde}{MDE}{Model-Driven Engineering}

\newacronym{ai}{AI}{Artificial Intelligence}

\newacronym{fdir}{FDIR}{Fault Detection, Isolation and Recovery}

\newacronym{ifm}{iFM}{Integrated Formal Methods}

\newacronym{ros}{ROS}{Robot Operating System}

\newacronym{amcl}{AMCL}{Adaptive Monte Carlo Localisation}

\newacronym{dl}{\text{\upshape\textsf{d{\kern-0.05em}L}}}{differential dynamic logic}

\newacronym{vm}{VM}{Virtual Machine}

\newacronym{bip}{BIP}{Behaviour Interaction Priority}

\newacronym{fol}{FOL}{First-Order Logic}

\newacronym{owl}{OWL}{Web Ontology Language}

\newacronym{ria}{RIA}{Restore Invariant Approach}

\newacronym{ocl}{OCL}{Object Constraint Language}

\newacronym{sil}{SIL}{Safety Integrity Level}

\newacronym{rcl}{RCL}{ROS Contract Language}

\newacronym{rv}{RV}{Runtime Verification}

\newacronym{rml}{RML}{Runtime Monitoring Language}
\usepackage{hyperref}
\usepackage{multirow}

\usepackage{textcomp}

\usepackage{tabularx}
\usepackage{wrapfig}
\usepackage{fretish} 
\usepackage{paralist}

\usepackage{tikz}
\usetikzlibrary{shapes.geometric, arrows}
\tikzstyle{fragment} = [rectangle, minimum width=2em, minimum height=2em, text centered, draw=black, fill={rgb:black,1;white,3}]
\tikzstyle{req} = [rectangle, minimum width=2em, minimum height=2em, text centered, draw=black]
\tikzstyle{arrow} = [thick,->,>=stealth]

\usepackage[]{todonotes}

\begin{document}

\title{Towards Refactoring FRETish Requirements\thanks{The authors thank Georgios Giantamidis, Stylianos Basagiannis, and Vassilios A. Tsachouridis (United Technologies Research Center, Ireland) for their collaboration in the requirements elicitation process; and Anastasia Mavridou (NASA Ames Research Center, USA) for her help with FRET. 
This research was financially supported by the European Union’s Horizon 2020 research and innovation programme under the VALU3S project (grant No 876852). This project is also funded by Enterprise Ireland (grant No IR20200054). The funders had no role in study design, data collection and analysis, decision to publish, or preparation of the manuscript.}}


\author{Marie Farrell \and Matt Luckcuck \and Ois\'{i}n Sheridan \and Rosemary Monahan}


\date{Department of Computer Science, Maynooth University, Ireland\\
\texttt{valu3s@mu.ie}}

\maketitle

\begin{abstract}

Like software, requirements evolve and change frequently during the development process. 
Refactoring is the process of reorganising software without changing its behaviour, to make it easier to understand and modify.
We propose refactoring for formalised requirements 
to reduce repetition in the requirement set so that they are easier to maintain as the system and requirements evolve. This work-in-progress paper describes our motivation for and initial approach to refactoring requirements in NASA's \gls{fret}. This work was directly triggered by our experience with an industrial aircraft engine software controller use case. In this paper, we reflect on the requirements that were obtained and, with a view to their maintainability, propose and outline functionality for refactoring \fretish{} requirements.

\end{abstract}

\glsresetall

\section{Introduction and Background}
\label{sec:intro}

Detailed requirements elicitation is an important step in the software development process. This often begins with a set of natural-language requirements, which then evolve as the project progresses, as additional functionality is added, and as bugs reveal unintended or unsafe system behaviour. For safety-critical systems, requirements can often be drawn from standards or regulator guidance, and verifying that the system's design and implementation preserve these requirements can be an integral part of securing approval to use the system. 

Formal methods can provide robust verification that gives developers and regulators the confidence that the system functions correctly and safely. However, natural-language requirements can be difficult to express in the logical formalisms that formal methods use. Tools such as NASA's \gls{fret} plug this gap by providing a structured natural requirements language (called \fretish{}) that has an underlying temporal logic semantics, which can be used directly as input to formal methods tools \cite{giannakopoulou2021automated}.


Through examining recent work \cite{deshpande2019data}, we see that sets of natural-language requirements can contain many similar requirements, as well as dependencies between requirements. 
This makes the necessary task of maintaining the requirements  tedious and error-prone, as the system and its requirements evolve. 

In software engineering, refactoring is the process of improving the structure of the software without altering its functionality \cite{fowler_refactoring_1999}. An example is using the \textsc{Extract Method} refactoring, which extracts a large piece of code into a method, to simplify and modularise the program. This is often used when the same functionality is repeated throughout the program. 
Here we investigate how to use refactoring to simplify and modularise requirements.

We view the maintenance of a requirements set to have similar benefits to the maintenance of software, namely that the requirements can be modified more easily with a reduced potential for human error. The notion of refactoring requirements is not new and has been previously explored in \cite{ramos_improving_2007}. 
Here, we introduce the idea for \gls{fret} 
through examining how refactoring can be applied to the \fretish{} requirements for an aircraft engine controller system.

Within the VALU3S project\footnote{\url{https://valu3s.eu/}}, we elicited and formalised requirements for an aircraft engine software controller use case with our industrial partner~\cite{luckcuck2021verifiable,farrell2021fretting}. We are now constructing formal models of the system to verify the requirements against, and generating verification conditions from the requirements. At this stage, it is important that our \fretish{} requirements are easy to maintain and update, should new or modified functionality be developed. As a result, we are devising an approach to refactoring these requirements to reduce repetition and aid the maintainabilty of the requirements set. 
We take inspiration from prior work on refactoring natural-language requirements \cite{ramos_improving_2007} and apply it to formal requirements with an additional step to check that the refactored requirement preserves the meaning of its unrefactored counterpart.



This work-in-progress paper explores how formalised \gls{fret} requirements can be refactored, and illustrates our refactoring process via our industrial, aerospace use case.

\section{Refactoring Requirements}
\label{sec:refactoringrequirements}

This section provides an overview and brief analysis of the requirements that we elicited for the aircraft engine software controller use case (originally presented in \cite{farrell2021fretting}) and describes our approach to refactoring them.

\subsection{Analysis: Aircraft Engine Controller Requirements}
\begin{table}[t]
\scalebox{0.9}{\begin{tabular}{|c|p{15cm}|}
\hline
\textbf{ID} & \textbf{Description}\\ \hline
UC5\_R\_1 & Under sensor faults, while tracking pilot commands, control objectives shall be satisfied (e.g., settling time, 
overshoot, and steady state error will be within predefined, acceptable limits) \\ \hline
UC5\_R\_2 & Under sensor faults, during regulation of nominal system operation (no change in pilot input), control objectives 
shall be satisfied (e.g., settling time, overshoot, and steady state error will be within predefined, acceptable limits) \\ \hline
UC5\_R\_3 & Under sensor faults, while tracking pilot commands, operating limit objectives shall be satisfied (e.g., respecting 
upper limit in shaft speed) \\ \hline
UC5\_R\_4 & Under sensor faults, during regulation of nominal system operation (no change in pilot input), operating limit 
objectives shall be satisfied (e.g., respecting upper limit in shaft speed) \\ \hline
\end{tabular}}
\caption{UC5\_R\_1--UC5\_R\_4 of the natural-language requirements for the aircraft engine controller. These 4 requirements are mainly concerned with continued operation of the controller in the presence of sensor faults \cite{farrell2021fretting}.}
\label{table: nlreqs}
\end{table}

Previously, we presented $14$ natural-language requirements for an industrial aircraft engine controller which we formalised using \gls{fret} \cite{farrell2021fretting}. Table \ref{table: nlreqs} contains the first 4 of these requirements, which were constructed independently by our industrial partner. It was clear to us from the outset that these requirements were repetitive, for example the phrase `\textit{Under sensor faults}' appears in several requirements ($4/14$ in total). 

To preserve traceability between the natural language requirements and their corresponding \fretish{} encodings we opted for a one-to-one mapping, where each natural-language requirement corresponds to one (parent) 
requirement in \fretish{}. \fretish{} requirements have the following structure and fields:

\centerline{\fretishComponents}
\noindent Here, \Scope{} and \Timing{} are optional. Users specify a \Condition{} under which a \Component{} shall satisfy a \Response{}. For example the \fretish{} encoding of UC5\_R\_1 is: 
\condition{if((sensorFaults)\&(trackingPilotCommands))} \component{Controller} \texttt{shall} \response{(controlObjectives)}. \textit{`Under sensor faults'} maps to the boolean \condition{sensorFaults}, and the other requirements (Table~\ref{table: nlreqs}) follow a similar structure. 

\begin{figure*}[t]
\hspace{50pt}
\scalebox{0.7}{
\begin{tikzpicture}[node distance=2em]

\node (TPC) [fragment, xshift = 50em]{\parbox[t][][t]{1cm}{\centering{F1}}};

\node (SF) [fragment, right of = TPC, xshift = 6em]{\parbox[t][][t]{1cm}{\centering F2}};

\node (CO) [fragment, right of = SF, xshift = 6em]{\parbox[t][][t]{1cm}{\centering F3}};

\node (OLO) [fragment, right of = CO, xshift = 6em]{\parbox[t][][t]{1cm}{\centering F4}};

\node (MF) [fragment, right of = OLO, xshift = 6em]{\parbox[t][][t]{1cm}{\centering F5}};

\node (LP) [fragment, right of = MF, xshift = 6em]{\parbox[t][][t]{1cm}{\centering F6}};

\node (RNO) [fragment, right of = LP, xshift = 6em]{\parbox[t][][t]{1cm}{\centering F7}};

\node (R1) [req, below of = TPC, yshift = -4em]{\parbox[t][][t]{1cm}{\centering{R1}}};

\node (R3) [req, below of = SF, yshift = -4em]{\parbox[t][][t]{1cm}{\centering{R3}}};

\node (R4) [req, below of = SF, xshift = 9em, yshift = -8em]{\parbox[t][][t]{1cm}{\centering{R4}}};

\node (R5) [req, below of = CO, yshift = -4em]{\parbox[t][][t]{1cm}{\centering{R5}}};

\node (R6) [req, below of = OLO, yshift = -4em]{\parbox[t][][t]{1cm}{\centering{R6}}};

\node (R7) [req, above of = OLO, yshift = 5em]{\parbox[t][][t]{1cm}{\centering{R7}}};

\node (R8) [req, below of = MF, xshift = 4em, yshift = -5em]{\parbox[t][][t]{1cm}{\centering{R8}}};

\node (R9) [req, above of = SF, yshift = 5em]{\parbox[t][][t]{1cm}{\centering{R9}}};

\node (R10) [req, below of = LP, yshift = -8em, xshift = -2em]{\parbox[t][][t]{1cm}{\centering{R10}}};

\node (R11) [req, above of = MF, yshift = 4em]{\parbox[t][][t]{1cm}{\centering{R11}}};

\node (R12) [req, above of = LP, yshift = 5em]{\parbox[t][][t]{1cm}{\centering{R12}}};

\node (R13) [req, above of = TPC, yshift = 4em]{\parbox[t][][t]{1cm}{\centering{R13}}};

\node (R14) [req, above of = RNO, yshift = 4em]{\parbox[t][][t]{1cm}{\centering{R14}}};

\node (R2) [req, above of = R7, yshift = 3em]{\parbox[t][][t]{1cm}{\centering{R2}}};

\draw [arrow ] (R1)  -- (TPC);
\draw [arrow] (R3)  -- (TPC);
\draw [arrow ] (R5)  -- (TPC);
\draw [arrow] (R7)  -- (TPC);
\draw [arrow] (R9)  -- (TPC);
\draw [arrow] (R11)  -- (TPC);
\draw [arrow] (R13)  -- (TPC);

\draw [arrow] (R1)  -- (SF);
\draw [arrow] (R2)  -- (SF);
\draw [arrow] (R3)  -- (SF);
\draw [arrow] (R4)  -- (SF);

\draw [arrow] (R1)  -- (CO);
\draw [arrow] (R2)  -- (CO);
\draw [arrow] (R5)  -- (CO);
\draw [arrow] (R6)  -- (CO);
\draw [arrow] (R9)  -- (CO);
\draw [arrow] (R10)  -- (CO);

\draw [arrow] (R3)  -- (OLO);
\draw [arrow] (R4)  -- (OLO);
\draw [arrow] (R7)  -- (OLO);
\draw [arrow] (R8)  -- (OLO);
\draw [arrow] (R11)  -- (OLO);
\draw [arrow] (R12)  -- (OLO);

\draw [arrow] (R5)  -- (MF);
\draw [arrow] (R6)  -- (MF);
\draw [arrow] (R7)  -- (MF);
\draw [arrow] (R8)  -- (MF);

\draw [arrow] (R9)  -- (LP);
\draw [arrow] (R10)  -- (LP);
\draw [arrow] (R11)  -- (LP);
\draw [arrow] (R12)  -- (LP);

\draw [arrow] (R2)  -- (RNO);
\draw [arrow] (R4)  -- (RNO);
\draw [arrow] (R6)  -- (RNO);
\draw [arrow] (R8)  -- (RNO);
\draw [arrow] (R10)  -- (RNO);
\draw [arrow] (R12)  -- (RNO);
\draw [arrow] (R14)  -- (RNO);

\node (TPCLabel) [fragment, below of = R1, yshift = -5em ]{\parbox[t][][t]{1cm}{\centering{F1}}};
\node (TPCdesc) [right of = TPCLabel, xshift = 7.5em]{\parbox[t][][t]{4.5cm}{Tracking \\ Pilot \\ Commands}};

\node (SFLabel) [fragment, right of = TPCdesc, xshift =-1em]{\parbox[t][][t]{1cm}{\centering{F2}}};
\node (SFdesc) [right of = SFLabel, xshift = 7.5em]{\parbox[t][][t]{4.5cm}{Sensor \\ Faults}};

\node (COLabel) [fragment, right of = SFdesc, xshift = -2em]{\parbox[t][][t]{1cm}{\centering{F3}}};
\node (COdesc) [right of = COLabel, xshift = 7.5em]{\parbox[t][][t]{4.5cm}{Control \\ Objectives}};

\node (OLOLabel) [fragment, right of = COdesc, xshift = -1em ]{\parbox[t][][t]{1cm}{\centering{F4}}};
\node (OLOdesc) [right of = OLOLabel, xshift = 7.5em]{\parbox[t][][t]{4.5cm}{Operating \\ Limit \\ Objectives}};

\node (MFLabel) [fragment, right of = OLOdesc, xshift = -1em ]{\parbox[t][][t]{1cm}{\centering{F5}}};
\node (MFdesc) [right of = MFLabel, xshift = 7.5em]{\parbox[t][][t]{4.5cm}{Mechanical \\ Fatigue}};

\node (LPLabel) [fragment, below of = TPCLabel, yshift = -2em ]{\parbox[t][][t]{1cm}{\centering{F6}}};
\node (LPdesc) [right of = LPLabel, xshift = 7.5em]{\parbox[t][][t]{4.5cm}{Low Probability \\ Hazardous Events}};

\node (RNOLabel) [fragment, right of = LPdesc, xshift = 2em] {\parbox[t][][t]{1cm}{\centering{F7}}};
\node (RNOdesc) [right of = RNOLabel, xshift = 7.5em]{\parbox[t][][t]{4.5cm}{Regulation of \\ Nominal Operation}};
\end{tikzpicture}

}
\caption{Dependency graph: arrows indicate a `depends on' relationship between requirements (white boxes) and fragments (grey boxes).}
\label{fig:dependency}
\vspace{-1em}
\end{figure*}
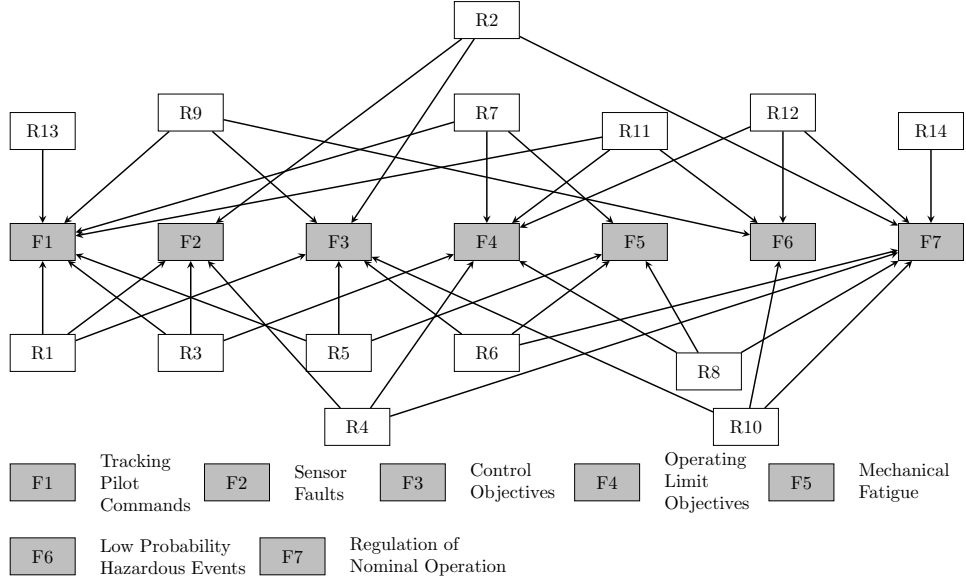

Since we adopted a one-to-one mapping, the repetition of \textit{`Under sensor faults'} is mirrored by the repetition of \condition{sensorFaults} in the \fretish{} requirements. We refer to these repeated pieces as requirement \textit{fragments}. We identified $7$ fragments in our $14$ abstract requirements, and each fragment was repeated in between $4$ and $7$ of the $14$ requirements. Fig.~\ref{fig:dependency} shows the dependencies between the requirements and specific fragments. 

Once the high-level requirements were encoded in \fretish, we elicited 28 detailed child requirements that expanded the definitions of the abstract terms in the 14 parent requirements \cite{farrell2021fretting}. UC5\_R\_1 has 3 child requirements (Table \ref{table: r1children}). Each of these contains the expanded, more detailed definition of \condition{sensorFaults}: \smallskip

\begin{minipage}{0.1\textwidth}
\end{minipage}
\begin{minipage}{0.9\textwidth}
\noindent\condition{(sensorValue(S) $>$ nominalValue + R) $|$ (sensorValue(S) $<$ nominalValue - R) $|$ (sensorValue(S) = null)}
\end{minipage}
\smallskip

\begin{table}[t]
\scalebox{0.9}{
\begin{tabular}{|c|p{15cm}|}
\hline
\textbf{ID}  &\textbf{FRETISH} \\ \hline
UC5\_R\_1.1  & \condition{when (diff(r(i),y(i)) $>$ E) if((sensorValue(S) $>$ nominalValue + R) $|$ (sensorValue(S) $<$ nominalValue - R) $|$ (sensorValue(S) = null) \& (pilotInput $=>$ setThrust = V2)  \& (observedThrust = V1))} \component{Controller} shall \timing{until (diff(r(i),y(i)) $<$ e)} \response{(settlingTime $>=$ 0) \& (settlingTime $<=$ settlingTimeMax) \& (observedThrust = V2)}\\\hline

UC5\_R\_1.2  & \condition{when (diff(r(i),y(i)) $>$ E) if((sensorValue(S) $>$ nominalValue + R) $|$ (sensorValue(S) $<$ nominalValue - R) $|$ (sensorValue(S) = null)\& (pilotInput $=>$ setThrust = V2) \& (observedThrust = V1))} \component{Controller} shall \timing{until (diff(r(i),y(i)) $<$ e)} \response{(overshoot $>=$ 0) \& (overshoot $<=$ overshootMax) \& (observedThrust = V2)}\\\hline

UC5\_R\_1.3  & \condition{when (diff(r(i),y(i)) $>$ E) if((sensorValue(S) $>$ nominalValue + R) $|$ (sensorValue(S) $<$ nominalValue - R) $|$ (sensorValue(S) = null)\& (pilotInput $=>$ setThrust = V2)\& (observedThrust = V1))} \component{Controller} shall \timing{until (diff(r(i),y(i)) $<$ e)} \response{(steadyStateError $>=$ 0) \& (steadyStateError $<=$ steadyStateErrorMax) \& (observedThrust = V2)} \\\hline
\end{tabular}}
\caption{Three distinct child requirements for UC5\_R\_1 capture the correct behaviour with respect to each of settling time, overshoot and steady state error.}
\label{table: r1children}
\end{table}

\noindent As expected, this repetition of definitions in the child requirements makes the requirements set more difficult to maintain, because changes to the definition of one fragment cause updates in multiple places. For example, if the definition of \condition{sensorFaults} were to change, as it did during the elicitation process, then $8$ of the $28$ child requirements would require updating. This process is time-consuming, tedious, and error prone. A better approach would be to update the definition of \condition{sensorFaults} in one place and avoid this duplication of effort. 

\condition{sensorFaults} corresponds to one detailed clause in each child requirement, but 
this was not the case for all  fragments. For example, \condition{trackingPilotCom\-mands} corresponds to a condition (\condition{when (diff(r(i),y(i))$>$E)}) and a timing constraint (\timing{until (diff(r(i),y(i))$<$e)}). 
An automatic approach to refactoring \fretish{} requirements would be even more helpful in similar situations where an abstract requirement corresponds to multiple detailed clauses.

Next, we outline our approach to refactoring \fretish{} requirements, taking inspiration from prior work on refactoring natural-language requirements.

\subsection{Refactoring Requirements}


We briefly show how we specialise the classical refactoring, \textsc{Extract Method}~\cite{fowler_refactoring_1999}, to requirements. \textsc{Extract Method} extracts code into a method, so that it 
can be called rather than copying code snippets.
Our specialisation is based on the \textsc{Extract Requirement} refactoring in \cite{ramos_improving_2007}; but with an extra step, facilitated by \gls{fret}'s automatic translation of requirements to temporal logic. 




We begin by creating a new requirement to contain the behaviour that we wish to extract. 
We then replace the extracted behaviour in the original requirement with a reference to the new one. Finally, we check that the restructuring has not altered the behaviour of the original requirement, and we propagate this change throughout the requirements set.

\textsc{Extract Requirement} allows us to define the \condition{sensorFaults} fragment in one place. Then, individual requirements essentially `call' the fragment in a similar way to method calls in object oriented programming languages. Supporting this `calling' capability in \gls{fret} is part of our current work. 

We chose \gls{fret} because it facilitates the formal verification that an implementation obeys its requirements. 
We intend to translate \fretish{} requirements into other formalisms for verification \cite{luckcuck2021verifiable}; so it is important that they are easy to maintain, if and when formal methods tools find problems in the system.

When 
refactoring the \fretish{} encodings of requirements we can formally verify that the refactoring preserves the semantics of the original requirements. The \gls{ltl} representation \gls{fret} generates enables us to perform the `compile and test' step that is included in software refactoring~\cite{fowler_refactoring_1999} but not previously addressed for refactoring natural-language requirements \cite{ramos_improving_2007}.



\section{Towards FRET-Supported Refactoring}

\gls{fret} does not currently support refactoring. This section outlines our initial investigations into how automatic refactoring functionality could be included.

\gls{fret} requirements are not aware of one another. For example, although a requirement might depend on 
another it cannot \textit{call the} other requirement in the way that a program can call a method. 
Requirements can be linked by 
a parent-child relationship but this is superficial at present, although it is useful from a user-perspective for maintaining traceability as requirements evolve.


We propose an additional requirement type, called \textsc{Fragment}, that can be called from other requirements. This will involve updating the \gls{fret} interface and will lead to minor modifications to the generation of \gls{ltl} specifications.
\gls{fret} uses an in-built bank of templates to generate the \gls{ltl} semantics for each requirement \cite{giannakopoulou2021automated}. Templates take the form: [\textit{$<$scope-option$>$, $<$condition-option$>$, $<$timing-option$>$}]. 
%
Since each \textsc{Fragment} will be a specialised requirement, each will produce a template. When generating the \gls{ltl} semantics for a requirement that references a \textsc{Fragment}, it will be necessary to combine the templates of the \textsc{Fragment} and the requirement to produce a complete template.

In general we think that combining templates can be achieved by taking the union of the $scope$, $condition$, and $timing$ fields (respectively). However, we can also see specific situations where this simple approach might fail; e.g., if we  combine two distinct timing options, should they be summed or should one take precedence (if so, which one)? We leave this investigation as future work.


Users should be able to refactor existing requirements \textit{and} create fragments from scratch. Refactoring existing requirements could be realised, similarly to refactoring code in Eclipse\footnote{Eclipse: \url{https://www.eclipse.org/ide/}.}, by selecting the part of a requirement to become a \textsc{Fragment} and selecting (from a context-menu) that it should be extracted. The \gls{fret} interface should also provide the option to `create \textsc{Fragment}'.

When refactoring existing requirements, \gls{fret} should check that the original and refactored requirements (including the extracted \textsc{Fragment}(s)) are equivalent. 
\gls{fret} already checks the equivalence of the past- and future-time \gls{ltl} for each requirement, this step performs a similar check between requirements.

\gls{fret} links to the CoCoSim \cite{bourbouh2020cocosim} and Copilot \cite{perez2020copilot} verification tools. The translations to these tools would now require an extra step to address refactored requirements. A naive approach would involve recombining the fragments, effectively `unrefactoring' the requirement. This would be hidden from the user, with the fully expanded requirement only appearing in the generated verification conditions. However, if the user wanted to edit the generated conditions, the original problems with repetition in the requirements would reappear. 

A more sophisticated approach would carry the refactoring relationship through to the generated conditions. For example, in CoCoSim, guarantees would be generated corresponding to the fragments, but investigating how these guarantees are combined whilst preserving the semantics of the requirement is future work.


\section{Conclusion}
\label{sec:conclusion}

This paper presents our work-in-progress on refactoring \gls{fret} requirements, which is directly motivated by our specification of an industrial aircraft engine controller use case. We demonstrated that repetition in natural-language requirements can cause difficulty when maintaining a set of corresponding formalised requirements, and presented an approach to refactoring requirements that extends an existing approach in the literature. We have also outlined how we intend to implement this in \gls{fret} as future work.

 Other \gls{fret} studies have not encountered such a strong need for refactoring \cite{bourbouh2021integrating,mavridou2020ten,farrell2022}. However,  these do not directly involve an industry partner throughout the requirements elicitation and formalisation process. Our study is unique, since it is the first published use of \gls{fret} in an industrial case study where development of the system is ongoing \cite{farrell2021fretting}. That said, recent \fretish{} requirements for a liquid mixer~\cite{mavridou2021partial} exhibit some repetition, so may benefit from our refactoring approach. Investigating refactoring for \gls{fret} in other use cases is an important avenue of future work.


\bibliographystyle{splncs04}
\bibliography{refactoring.bib}

\end{document}